\documentclass[prl, linenumbers, superscriptaddress, twocolumn]{revtex4-1} 
\usepackage[utf8]{inputenc}
\usepackage{mathrsfs}
\pdfoutput=1
\usepackage{color}
\usepackage{graphicx}
\usepackage{epstopdf}
\usepackage{amsmath}
\usepackage{amsfonts}
\usepackage{amssymb}
\usepackage{latexsym}
\usepackage{subfigure}
\newcommand{\cH}{\mathcal H}

\newcommand{\cM}{\mathcal M}
\newcommand{\cI}{\mathcal I}

\newcommand{\bbI}{\mathbb I}

\newcommand{\bk}{\mathbf k}

\definecolor{BrickRed}{cmyk}{0,0.89,0.94,0.28}%%%PANTONE 1805
\definecolor{MidnightBlue}{cmyk}{0.98,0.13,0,0.43}%%%PANTONE 302
\definecolor{DarkGreen}{rgb}{0,0.7,0.1}

\definecolor{RedViolet}{cmyk}{0.07,0.90,0,0.34}%%%PANTONE 234
\definecolor{SeaGreen}{cmyk}{0.69,0,0.50,0}%%%PANTONE 3268
\definecolor{FireOrange}{rgb}{1.,.294,.247}

\begin{document}
\title{Bulk Dirac points in distorted spinels}
\author{Julia A. Steinberg}
\affiliation{The Makineni Theoretical Laboratories, Department of Chemistry, 
University of Pennsylvania,
Philadelphia, Pennsylvania 19104-6323, USA}
\affiliation{Department of Physics and Astronomy, University of Pennsylvania, 
Philadelphia, Pennsylvania 19104-6396, USA}
\author{Steve M. Young}
\affiliation{The Makineni Theoretical Laboratories, Department of Chemistry, 
University of Pennsylvania,
Philadelphia, Pennsylvania 19104-6323, USA}
\author{Saad Zaheer}
\affiliation{Department of Physics and Astronomy, University of Pennsylvania, 
Philadelphia, Pennsylvania 19104-6396, USA}
\author{C. L. Kane}
\affiliation{Department of Physics and Astronomy, University of Pennsylvania, 
Philadelphia, Pennsylvania 19104-6396, USA}
\author{E. J. Mele}
\affiliation{Department of Physics and Astronomy, University of Pennsylvania, 
Philadelphia, Pennsylvania 19104-6396, USA}
\author{Andrew M. Rappe}
\affiliation{The Makineni Theoretical Laboratories, Department of Chemistry, 
University of Pennsylvania,
Philadelphia, Pennsylvania 19104-6323, USA}
\begin{abstract}
We report on a Dirac-like Fermi 
surface in three-dimensional bulk materials in a distorted spinel structure on 
the basis of density functional theory (DFT) as well as tight-binding theory. 
The four examples we provide in this paper are $\rm BiZnSiO_{4}, BiCaSiO_{4}, 
BiMgSiO_{4}, and~BiAlInO_{4}$. A necessary characteristic of these structures is 
that they contain a Bi lattice which forms a hierarchy of chain-like 
substructures, with consequences for both fundamental understanding and 
materials design.
\end{abstract}
\maketitle
Following the discovery of topological insulators, there has been considerable interest in studying semimetallic phases that exist at the phase transition between a topological and a trivial insulator. One such example is graphene, which has two Dirac points at its Fermi surface. A Dirac point is characterized by four degenerate states that disperse linearly with momentum around a single point $\bk$ in the Brillouin zone. The resulting low energy theory is pseudorelativistic, and it is responsible for many of the interesting properties of graphene  \cite{RevModPhys.81.109}. In a previous communication, we described such Dirac points occurring as symmetry-protected fourfold degeneracies in three dimensional crystal systems. Such a Dirac point was encountered first in a tight binding model of $s$-states on the diamond lattice \cite{PhysRevLett.98.106803}. We showed that the occurrence of this Dirac point is a feature of the symmetry of diamond, which occurs in spacegroup 227. While no realistic system of atoms in a primitive 
diamond lattice produces this feature, through first-principles calculations we 
found that substituting bismuth for silicon in $\beta$-cristobalite, also in 
space group 227, elevates a candidate Dirac point degeneracy to the Fermi 
level, making BiO$_{2}$ the first realistic Dirac semimetal 
proposed~\cite{PhysRevLett.108.140405}. While this material is predicted to be 
metastable, it is highly unfavorable thermodynamically. This led us to consider 
alternative crystal structures and symmetries. 3D Dirac points have also been predicted to exist at the phase transition between a topological and a normal insulator when inversion symmetry is present~\cite{Murakami07p356,Youngp11085106}. If either inversion or time reversal symmetry is broken at the transition, a Dirac point separates into Weyl points which have been shown to exist in Refs.~\cite{Vishwanath11p205101,Burkov11p127205,Balents11arXiv}. Here we find, on the basis of density functional theory and tight binding calculations, that a family of three dimensional materials in a distorted spinel structure support robust Dirac points in their bulk electronic spectra. A common characteristic of these materials is the presence of a hierarchy of chain-like substructures created by the Bismuth atoms in the $A$ site of the crystal structure.

The spinel structure hosts a family of chalcogenides which have the general 
formula $AB_{2}X_{4}$. $A$ and $B$ are cations that are coordinated by anions 
of species $X$, which may be O, S, Se, or Te. The $A$ sites are tetrahedrally 
coordinated in a diamond lattice, with the octahedrally coordinated $B$ sites 
in the interstices~\cite{doi:10.1021/ja00365a019}. Most known spinels are 
insulators with band gaps of a few eV; this natural tendency toward insulating 
behavior makes the spinel structure a prime candidate to host a material with a 
distinct Dirac point degeneracy formed by bands that do not elsewhere cross the 
Fermi level. By contrast, though the laves structure contains the required 
symmetry for a Dirac point, materials in this structure tend to be metallic and 
our attempts to engineer Dirac semimetals in this structure were plagued by 
additional band crossings at the Fermi level.

\begin{figure}
\centering
{
\subfigure[227]{\includegraphics[width=0.48\textwidth,bb=0 0 1000 
750]{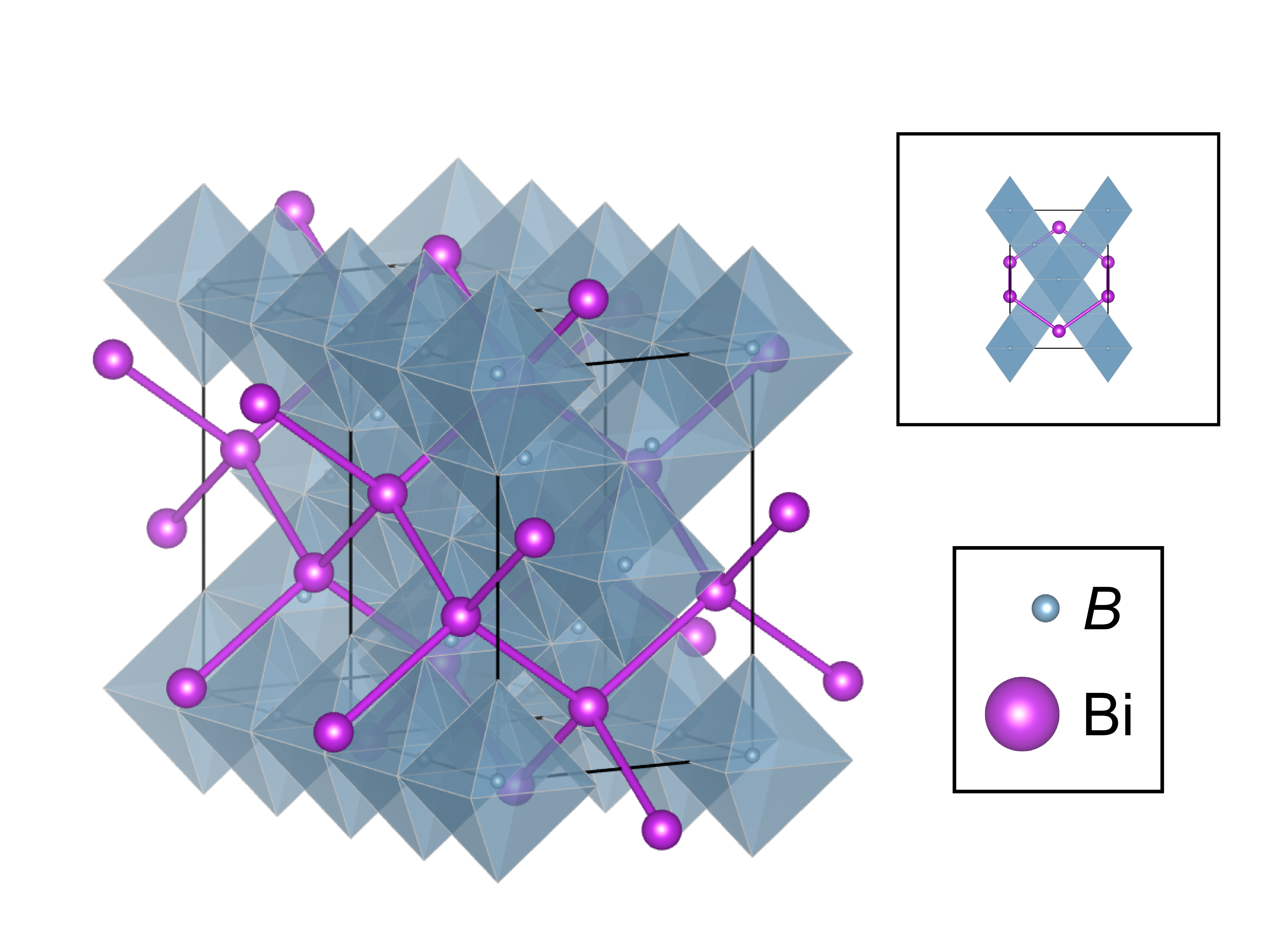} \label{fig:227}}
\subfigure[74]{\includegraphics[width=0.48\textwidth,bb=0 0 1000 
750]{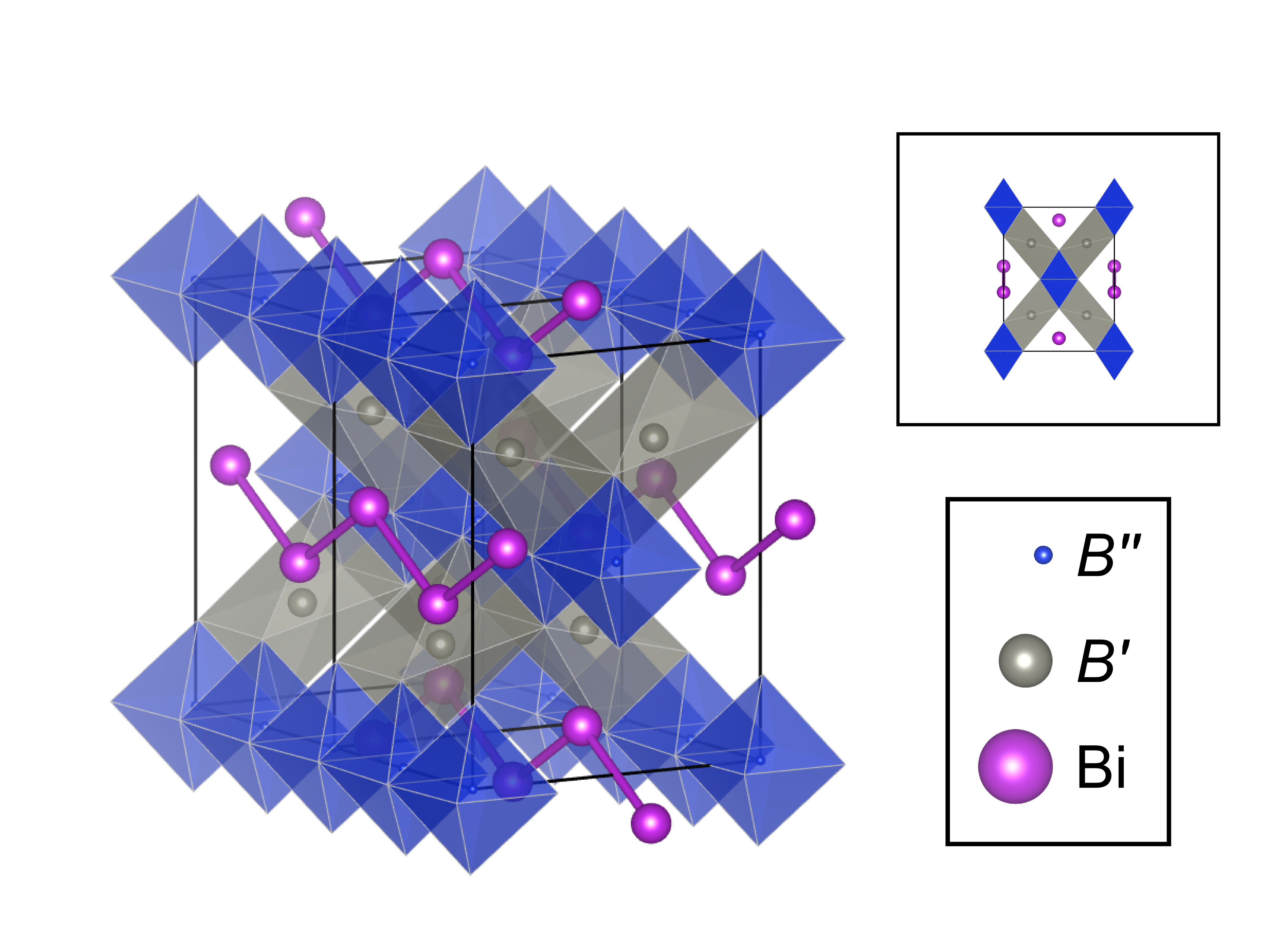} \label{fig:74}}
}
\caption{The~\subref{fig:227} spinel and~\subref{fig:74} distorted spinel 
structures. We note that the high-symmetry spinel conventional unit cell is 
cubic; here we choose a reduced supercell allowing for direct comparison with 
the distorted structure. The $A$ site, populated here by bismuth, is surrounded 
by a tetrahedral oxygen cage (not shown for clarity) and the $B$-type sites are 
surrounded by octahedral oxygen cages. In the high-symmetry spinel all Bi 
neighbors are equidistant, and the bismuth atoms form a diamond lattice.  Upon 
distortion, bonds lying in the $(0,1,0)$ plane elongate, and the structure can 
be described as coupled, close-packed, corrugated chains running parallel to 
$\left[ 0,1,0\right]$.  }
\label{spinelstructs}
\end{figure}

\begin{figure}
\centering
{
\subfigure[BiZnSiO$_{4}$]{\includegraphics[width=0.35\textwidth]{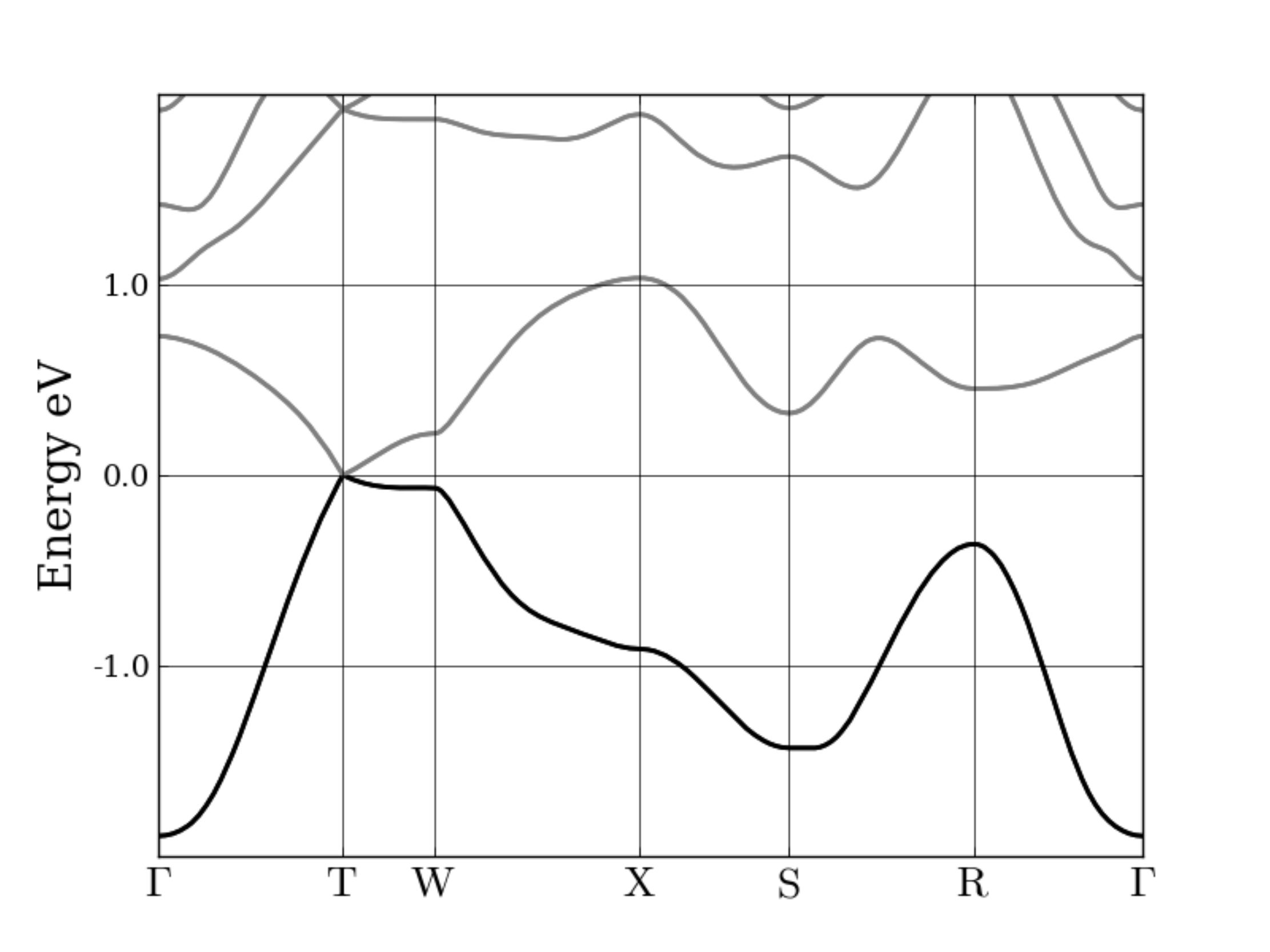} \label{fig:bandstructure1}}
\subfigure[BiCaSiO$_{4}$]{\includegraphics[width=0.35\textwidth]{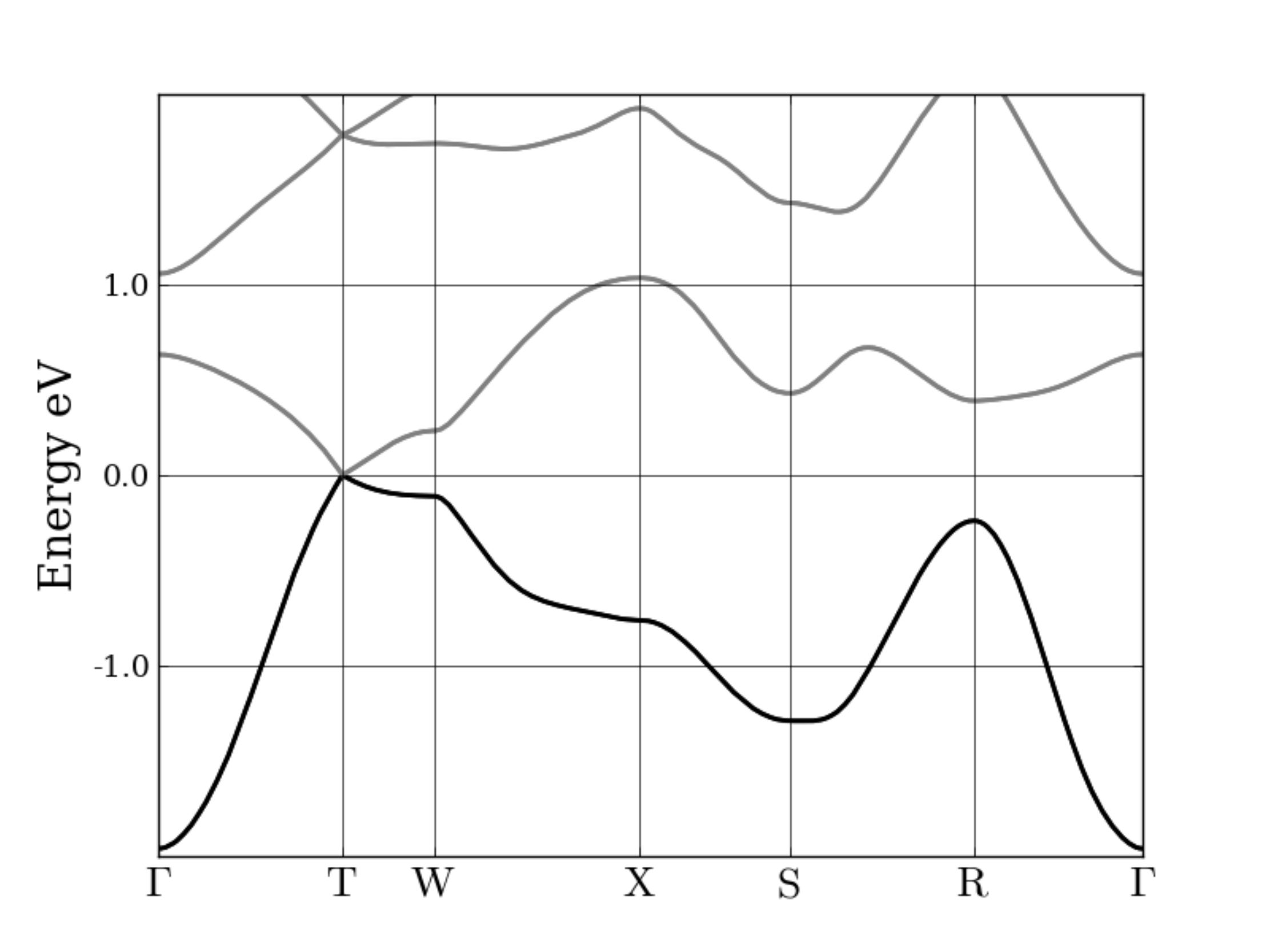} \label{fig:bandstructure2}}
\subfigure[BiAlInO$_{4}$]{\includegraphics[width=0.35\textwidth]{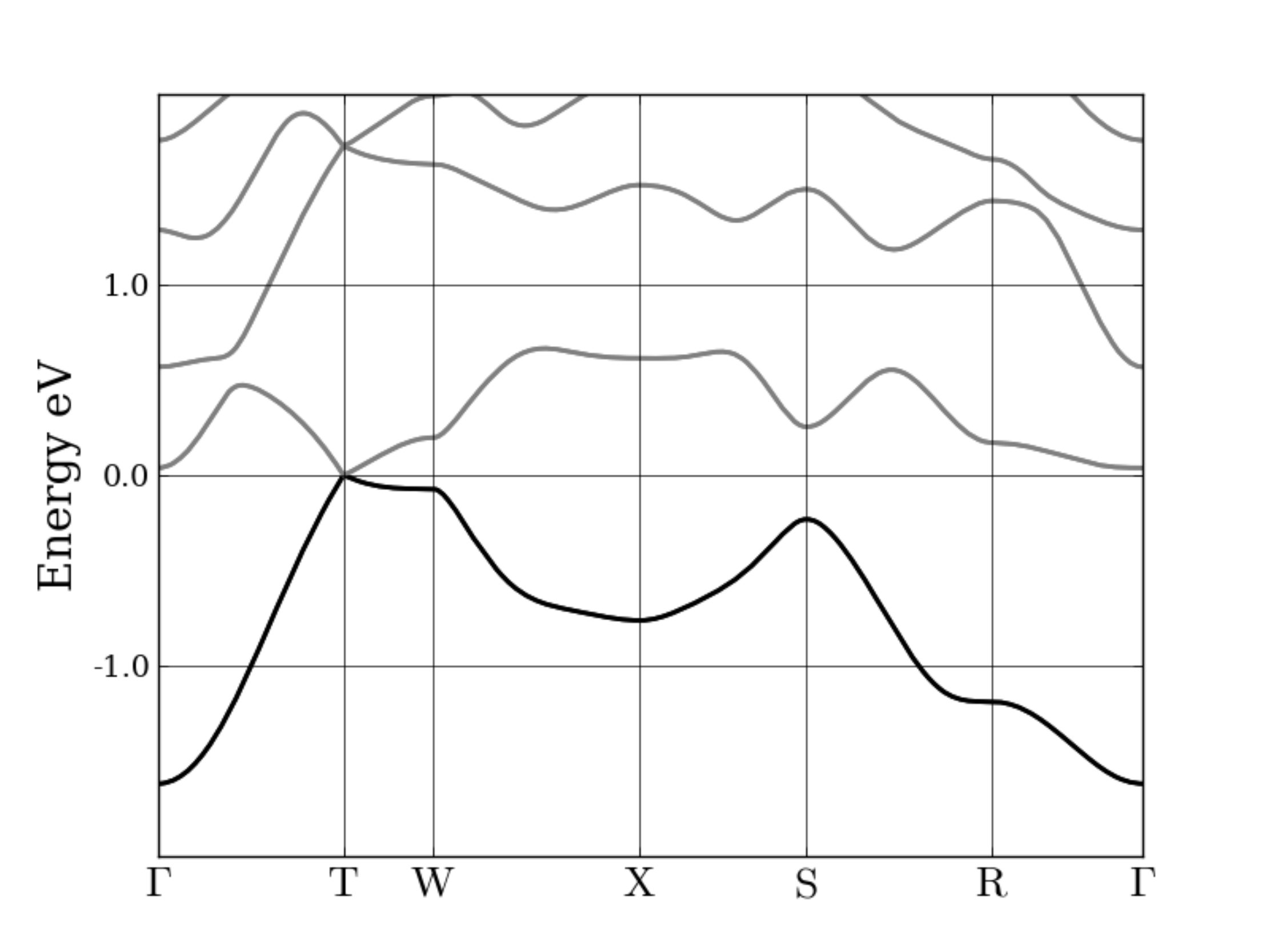} \label{fig:bandstructure3}}
\subfigure[BiMgSiO$_{4}$]{\includegraphics[width=0.35\textwidth]{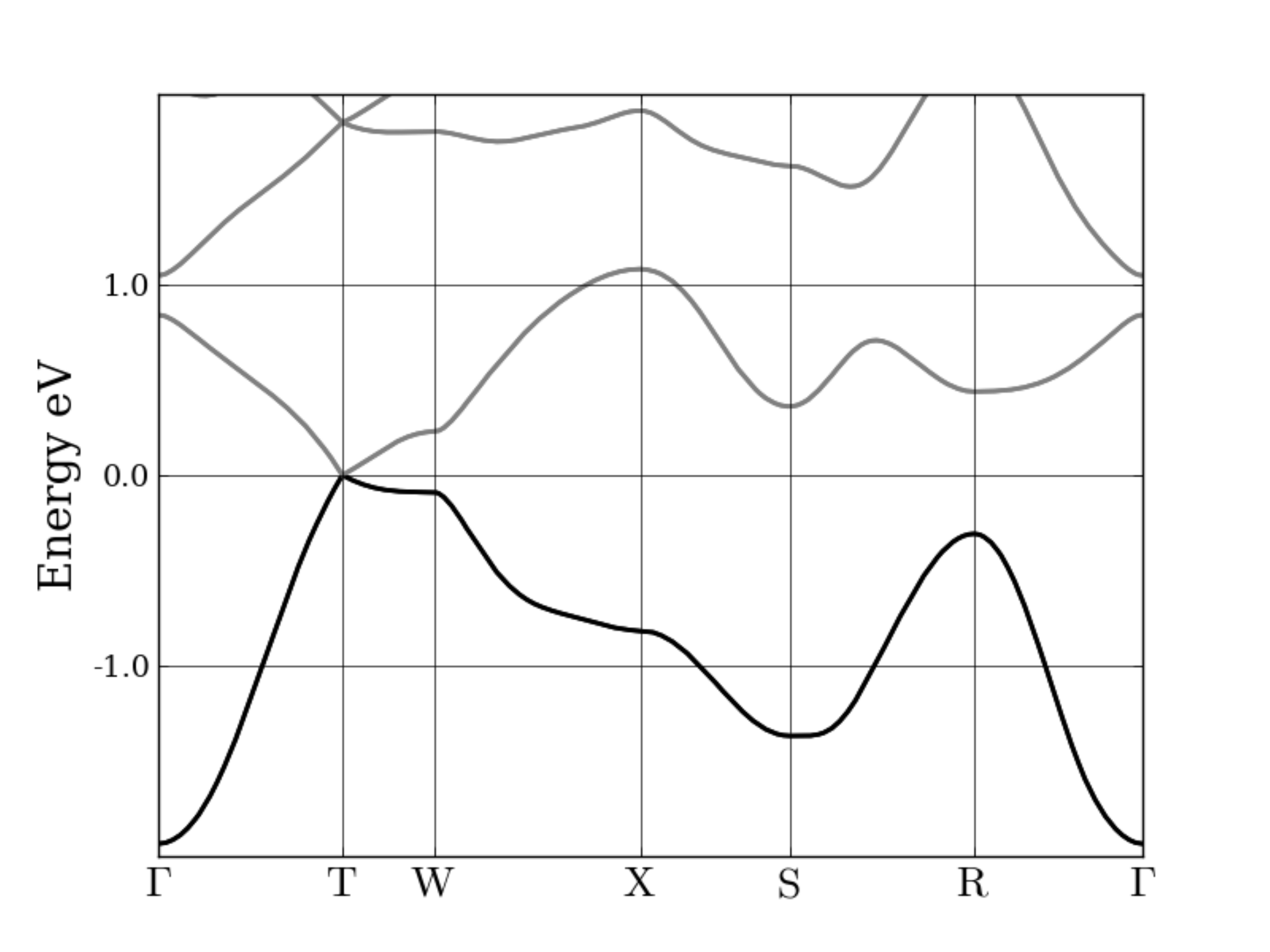} \label{fig:bandstructure4}}}
\caption{The bandstructures of BiZnSiO$_{4}$ (\ref{fig:bandstructure1}), 
BiCaSiO$_{4}$ (\ref{fig:bandstructure2}), 
BiAlInO$_{4}$(\ref{fig:bandstructure3}), and BiZnSiO$_{4}$ 
(\ref{fig:bandstructure4}) in the distorted spinel structure all contain a 
Dirac point at $\rm T$ that is completely split along the line in the Brillouin 
zone from $\rm T$ to $\rm W$ due to the spin orbit interaction between the 
bismuth sites in the lattice.}
\label{bands}
\end{figure}

BiAl$_{2}$O$_{4}$ and BiSc$_{2}$O$_{4}$ in the spinel structure exhibit Dirac 
points at the X points in the BZ.  However, these materials are not stable and 
spontaneously break symmetry.  Fortunately, one of the child symmetry groups is 
space group 74, which our crystallographic symmetry 
criteria~\cite{PhysRevLett.108.140405} admit as a potential host for a Dirac 
semimetal. As shown in Fig.~\ref{spinelstructs}, the cubic unit cell distorts 
to an orthorhombic cell and the bismuth atoms shift from their previous 
locations along one of the axes of twofold rotation symmetry in the diamond 
lattice. This symmetry breaking also distinguishes the two $B$ atoms of the 
formula unit, which we label $B'$ and $B''$, so that the composition becomes 
Bi$B'B''O_4$.  While this new space group has much lower symmetry, it inherits 
the non-symmorphic symmetry of diamond and retains one the three Dirac points 
among its representations. This Dirac point occurs at the point T in the 
Brillouin (BZ) and, as in diamond, there is a 
gap at 
W determined by the strength of the spin-orbit interaction.  The effects of 
reduced symmetry are manifest in the elongation of the octahedral cages of the 
$B'$ site. This allows the diamond lattice of bismuth to separate into 
distinct, parallel zig-zag chains.  
Through first principles DFT calculations we have determined that 
MgBiSiO$_{4}$, AlBiInO$_{4}$, ZnBiSiO$_{4}$, and CaBiSiO$_{4}$ are each 
metastable and exhibit Dirac point degeneracies at T with no other band 
crossings at the Fermi level (Fig.~\ref{bands}).  Both the crystal and 
electronic structures are very similar for these materials, and in the 
following we take BiZnSiO$_4$ to be representative of all four. For these 
electronic structure and atomic relaxation calculations, we used the 
plane wave density functional theory package QUANTUM ESPRESSO 
\cite{quantumespresso} and designed nonlocal pseudopotentials 
\cite{PhysRevB.41.1227, PhysRevB.59.12471} with spin-orbit interaction 
generated 
by OPIUM.  For all calculations an energy cutoff of 50~Ry and k-point grids of 
$8\times 8\times 8$ for the primitive cell were used. 

\begin{figure}
%\centering
{
\subfigure[]{\includegraphics[height=2in]{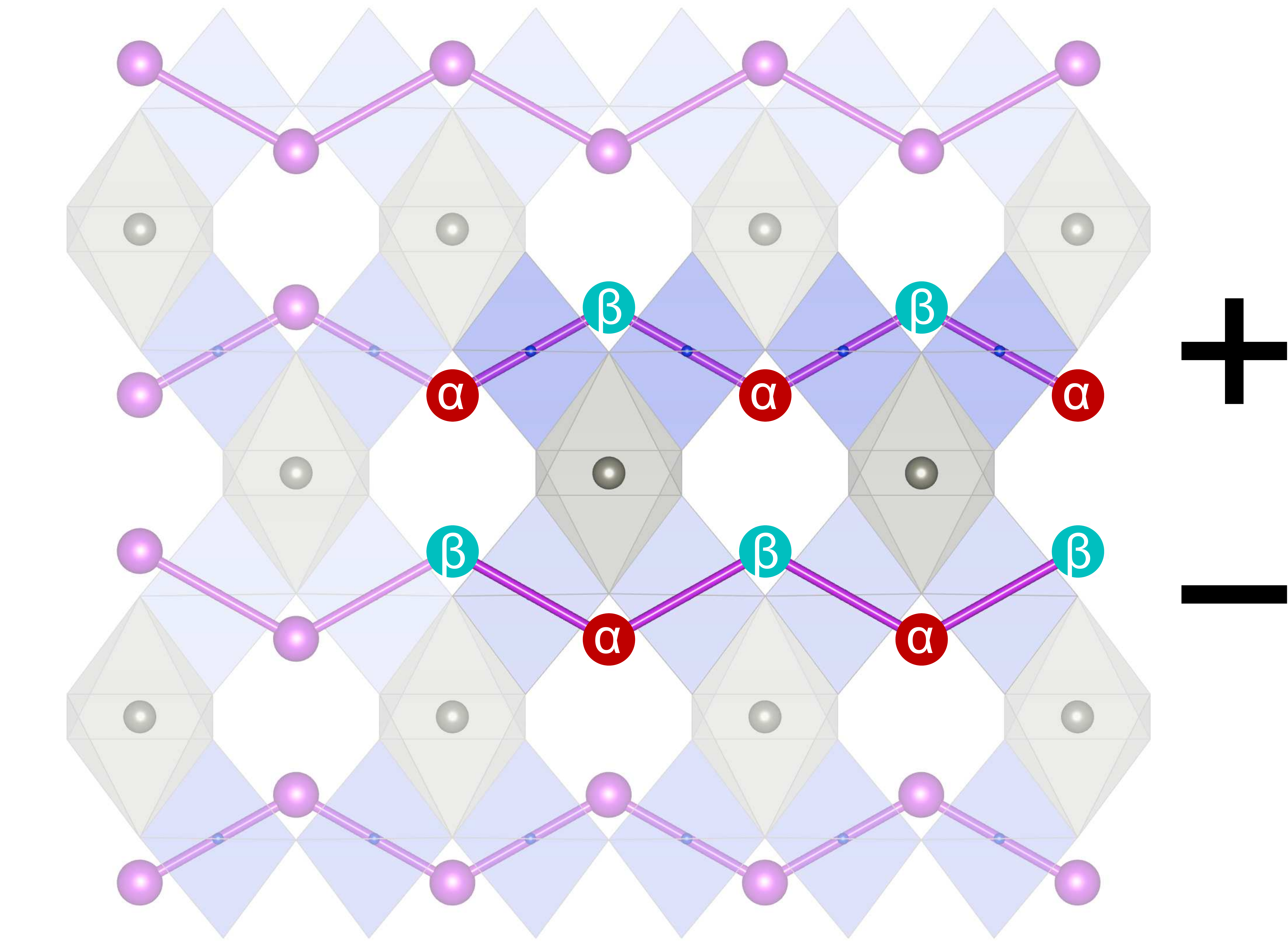} \label{chain1}}
\subfigure[]{\includegraphics[height=2in]{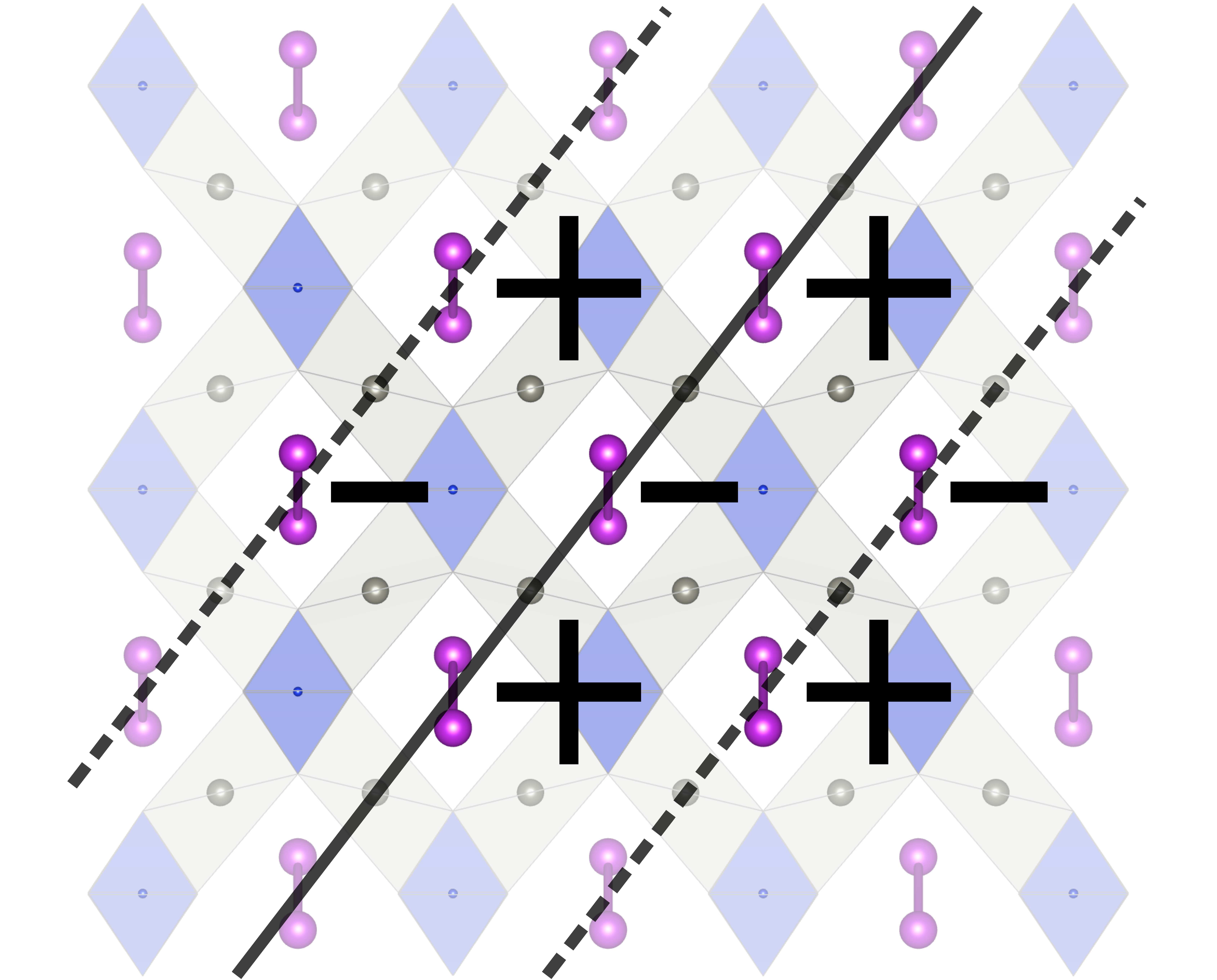} \label{chain2}}}
\caption{The distorted spinel structure with the bismuth network annotated. 
$B''$ cages are blue, $B'$ cages are gray, and the bismuth atoms are in purple. 
 The bismuth chains~\subref{chain1} are composed of bismuth atoms with 
alternating bond directions, labeled $\alpha$ and $\beta$, forming chains with 
distinct orientations from their nearest neighbors, signified by $+$ and $-$.  
Running parallel to $\left[  0, 1, 0\right]$, the alternatingly oriented chains 
form their own chain like structure in the $\left[1,0,1\right]$ direction, so 
that each chain is now an object coupling to its neighbors in 1D.  The 
resulting parallel planes of alternating sense (denoted by solid and dashed 
lines), constructed from the sheets of coupled chains, themselves couple with 
one another in the $\left[1,0,-1\right]$ direction.  Thus, at each level of 
structure the elements (atoms, chains, and planes) alternate in orientation 
along a particular direction, resulting in a four-fold representation for a 
point on the BZ surface 
corresponding to half a reciprocal lattice vector for each of those directions, 
and a Dirac point at the Fermi level for half-filling. }
\label{chains}
\end{figure}
\begin{figure}
\centering
{
\subfigure[]{\includegraphics[width=0.4\textwidth]{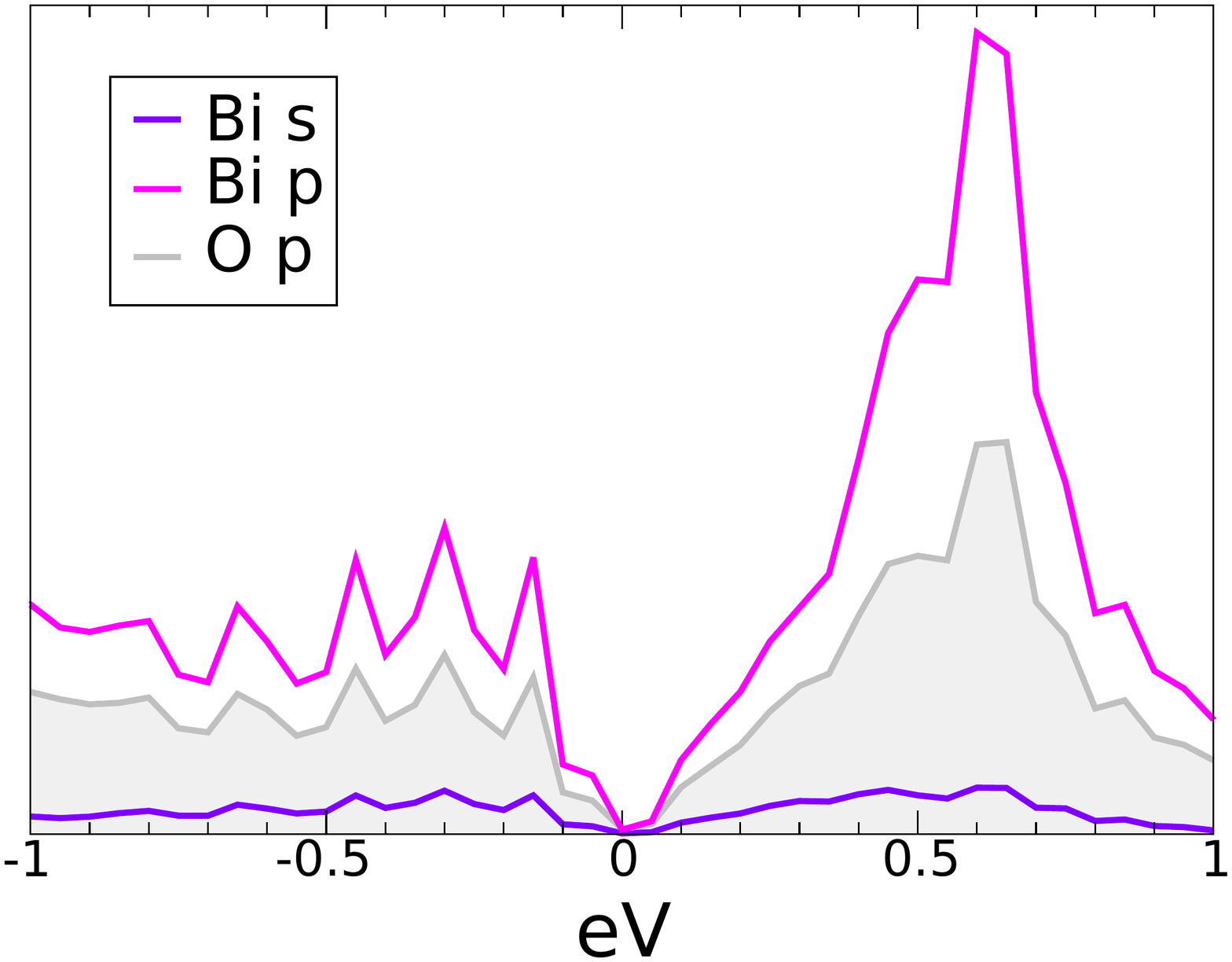} 
\label{fig:pdos}}
\subfigure[]{\includegraphics[height=2in,bb=0 0 875 1333]{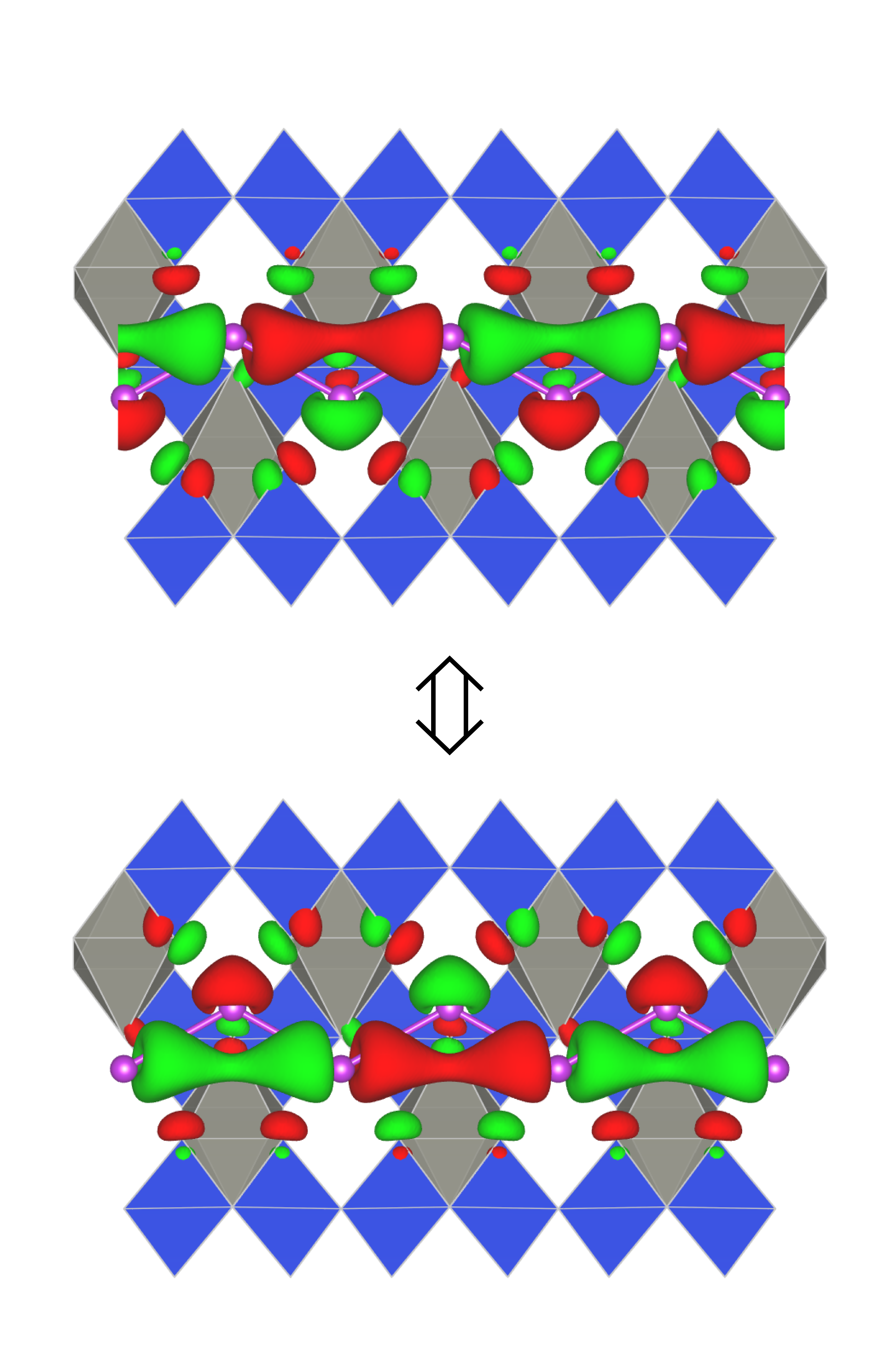} 
\label{fig:wavefunctions}}
\subfigure[]{\includegraphics[height=2in,bb=0 -100 1750 1392]{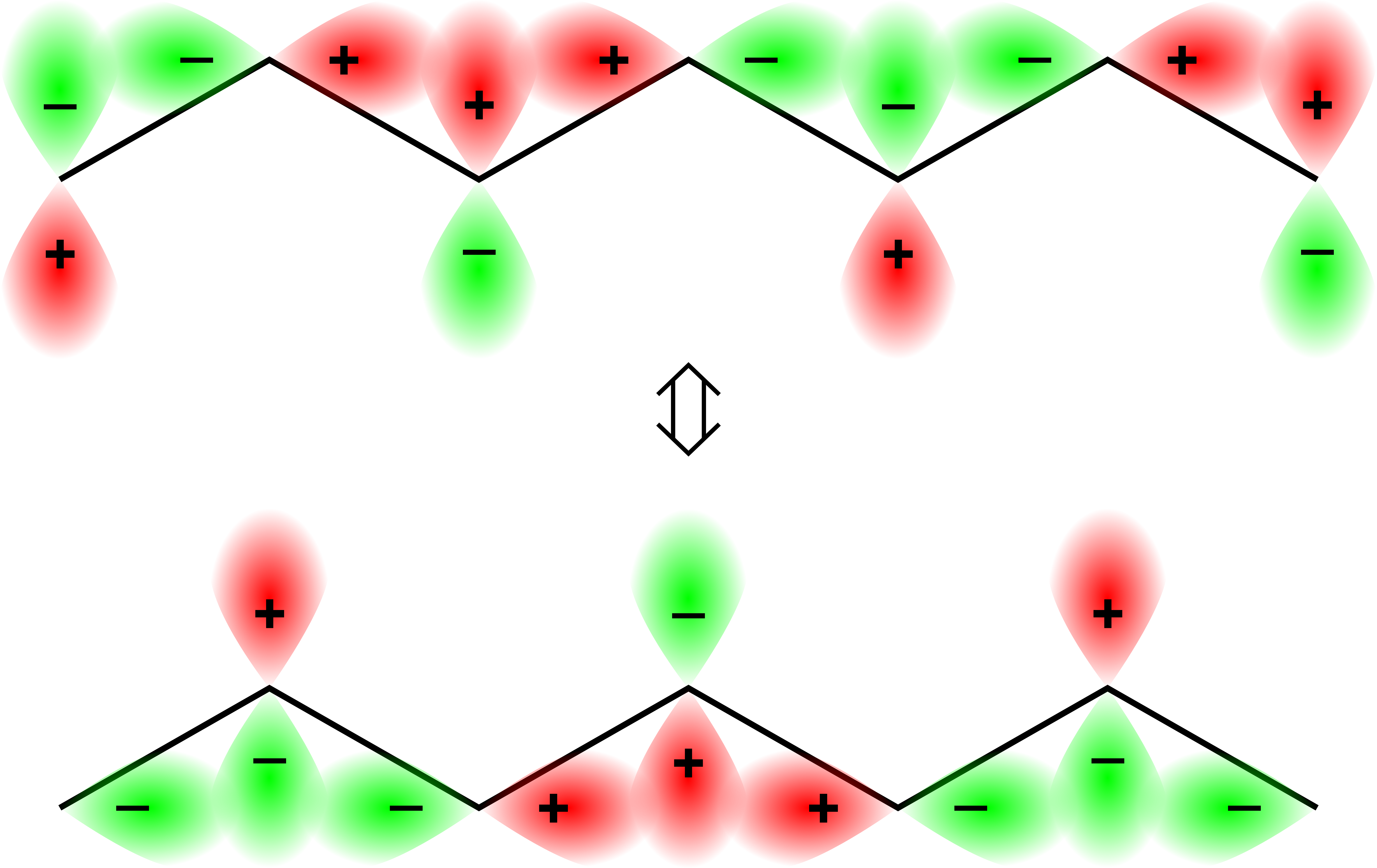} 
\label{fig:bismuthpchain}}

}
\label{electronic}
\caption{A plot of the projected density of states near the Dirac point in the 
ZnBiSiO$_{4}$ structure (\ref{fig:pdos}) shows that the states near the Dirac 
point are primarily bismuth $p$-states, with all states vanishing at the Fermi 
energy. The angular portion of the wavefunctions (ie, without the spin degree 
of freedom) of the ZnBiSiO$_{4}$ structure at $\rm T$ are shown 
in~\subref{fig:wavefunctions}, with a cartoon depiction in 
\subref{fig:bismuthpchain} emphasizing their $p$-like nature. The neighboring 
bismuth sites interact in a way that can be depicted as chains of bismuth 
$p$-orbitals running through an insulating structure, in which individual 
chains are analogous to polyacetylene; it is clear that the nonsymmorphic 
operation of inversion coupled to translation between A sites results in 
equivalent but different state, creating a twofold degeneracy at the Dirac 
point, which then becomes fourfold due to the spin degree of freedom. Breaking 
inversion symmetry in this structure is analogous to having 
inequivalent coupling constants between bismuth sites and single and double 
bonds in polyacetylene, and will lift the degeneracy, gapping the system.}
\end{figure}

The states near the Fermi surface are dominated by $p$-like states on the 
bismuth atoms, as revealed by the angular-momentum-projected density of states 
in Fig.~\ref{fig:pdos}), suggesting that each bismuth possesses a pair of 5$s$ 
electrons and an unpaired 5$p$-electron.  This contrasts with both the 
Fu-Kane-Mele tight-binding model and BiO$_2$, where the electronic character of 
the Dirac point derives from an unpaired 5$s$-electron.  The presence of an 
unpaired electron is required by symmetry considerations. Nonsymmorphic space 
groups have at least one sublattice degree of freedom, and at the k-points 
hosting Dirac points, the only representations are fourfold.  Thus, for a 
fourfold degeneracy to be bisected by the Fermi level there must be an odd 
number of electrons per formula unit. 
This symmetry constraint signifies physics being driven by these unpaired 
electrons. In Fig.~\ref{fig:wavefunctions} the Bloch states (excluding the spin 
degree of freedom) of the Dirac point degeneracy are shown, with a cartoon 
representation in Fig.~\ref{fig:bismuthpchain} added for clarity. The character 
of these states confirms that the unpaired $p$-electrons give rise to the 
observed Dirac point physics: the two states are related by the symmetry 
between the two bismuth sublattices that makes the space group nonsymmorphic.  
Additionally, the system must lie at a critical point that lies between the two 
configurations where bismuth atoms pair into dimers and is protected by the 
sublattice symmetry in all three directions; otherwise the interaction between 
the zig-zag chains could gap the system. This symmetry, which prevents the 
unpaired electrons from forming bonds in either direction, only belongs to the 
little group at T; elsewhere it is absent and the degeneracy between the two 
states is lifted.
These chains are reminiscent of polyacetylene, which is characterized by a 
pattern of alternating single and double bonds along carbon sites leading to a 
two-fold degenerate ground state. This suggests that the individual chains of 
bismuth sites behave like coupled one-dimensional metal wires running through 
an otherwise insulating structure. An isolated chain of this kind would be 
described by the Su-Schrieffer-Heeger~\cite{PhysRevLett.42.1698} model and 
would suffer from the Peierls instability intrinsic to the half-filled state, 
breaking the Dirac point.  However, the coupling between the chains requires a 
three-dimensional model, and the stability against dimerization depends on 
microscopic details.   First, upon dimerization, the oxidation state of bismuth 
becomes defined as 2+.  This is highly unfavorable for bismuth, which is known 
to prefer oxidation states of 3+ or 5+.  Second, the other cations are small, 
encouraging a more closely packed lattice and increasing the favorability of 
the delocalized, 
metallic character of the Dirac point. Thus the critical point, where the 
oxidation state of bismuth is formally undefined and the unpaired electrons are 
delocalized, is locally stable.

Focusing now on the bismuth lattice, Figure~\ref{chains} shows an annotated 
structure of a distorted spinel. The bismuth atoms form a chain like structure 
going into the plane, and adjacent chains have a different ordering of the 2 
type bismuth atoms. Fig.~\ref{chain1} illustrates chains of $\alpha$- and 
$\beta$-type Bi atoms along the $y$-axis, while Fig.~\ref{chain2} illustrates 
that adjacent atoms along the $x$ and $z$-axes form an additional chain-like 
structure. Therefore the Dirac point at T can be understood to be arising from 
three levels of  chain-like structures, resulting in a Dirac point that is 
protected by the sublattice symmetry of space group 74. 

We may model the low-energy theory of distorted spinels by a tight binding 
model of $p$-states on the $\rm Bi$ atoms. The two bismuth atoms in each unit 
cell each have $p_x,p_y,$ and $p_z$ orbitals. We distinguish between the two 
bismuth sites (and associated sublattices) with the labels A and B. There are 
four bonds possible for each site; these are ${\bf d}^{1\pm}=\left(\pm a/2, 0, 
[1-2\gamma]c\right)$ and ${\bf d}^{2\pm}=\left(0, \pm b/2, 2\gamma c\right)$, 
where $a$, $b$, and $c$ are the lengths of the orthorhombic lattice vectors, 
and $\gamma$ describes an internal distortion; when $a=b=c/\sqrt{2}$ and 
$\gamma=1/8$, the lattice becomes diamond.  Excluding spin for the moment, the 
tight binding Hamiltonian becomes
\begin{eqnarray}
\cH_{\rm tb} = & \sum_{<ij>} c^{\dagger}_{i,\alpha} c_{j,\beta} 
[\left(t_\sigma-t_\pi\right)\left({\bf \alpha}\cdot {\bf 
d}_{ij}\right)\left(\beta\cdot {\bf d}_{ij}\right)\\ \nonumber &+t_\pi\left(\alpha\cdot 
\beta\right)]\label{eq:tightbinding}
\end{eqnarray}
%The bismuth sites are indexed by $i$ and $j$, and the sets of $p$ orbital 
orientations for $\alpha$-sites and $\beta$-sites are labeled by $\alpha$ and 
$\beta$, which may be the unit vectors $\hat{x}$, $\hat{y}$, and $\hat{z}$.
The bismuth sites are indexed by i and j, and the sets of p orbital 
orientations for site i are labeled by $\alpha$ whereas those for site j are 
labeled by $\beta$, which may be the unit vectors $\hat{x}$, $\hat{y}$, and 
$\hat{z}$. $t_\sigma$ and $t_\pi$ are phenomenological coupling parameters for 
the $\sigma$ and $\pi$ character of the $p$-$p$ bonds, and ${\bf d}_{ij}$ is 
one of the aforementioned bond vectors that connect sites $i$ and $j$.

The resulting Hamiltonian produces three pairs of bands, each with a degeneracy 
from T to W, with bonding and anti-bonding pairs split off below and above the 
middle non-bonding pair by energy proportionate to $\left| 
t_\sigma-t_\pi\right|$.  Since each bismuth contributes a single electron, we 
expect the lowest, bonding pair of bands to be half-filled.  By inspection we 
find that these bands at T are dominated by the $p$ orbitals in the plane of 
the chain, corroborating our first principles results, and confirming that the 
crucial physics is due to the bismuth lattice.  Introducing a spin-orbit term 
of the form 
\begin{eqnarray}
\cH_{\rm so}= \sum_{<<ij>>, ss', \alpha\beta}i\lambda_{i\alpha j\beta} {\bf 
d}^1_{ij}\times {\bf d}^2_{ij}\cdot \vec{\sigma}_{ss'} c^{\dagger}_{i,\alpha 
s}c_{j,\beta,s'}. 
\end{eqnarray}
${\bf d}^1_{ij}$ and ${\bf d}^2_{ij}$ are nearest neighbor bond vectors that 
connect sites $i$ and $j$ on the same sublattice, we find that the degeneracies 
at W are split, allowing Dirac points at T, but the effect is not strong enough 
to mix the three sets of bands with one another, again in agreement with the 
first-principles calculations.
Removing the distortion in space group 74 to restore the diamond lattice in 
Eq.~\eqref{eq:tightbinding} provides the three Dirac points originally known to 
exist at the X-points in diamond. Thus, high symmetry diamond exists as a 
critical point between the single Dirac-point phases allowed by the three 
directions in which the symmetry may be reduced to space group 74.  This 
provides a simple understanding of how the Dirac point in diamond is connected 
to the Dirac point in space group 74.

To evaluate the possibility of synthesizing a Dirac semimetal in the 
laboratory, we calculated the energy difference associated with synthesizing  
ZnBiSiO$_{4}$ from zinc silicate, bismuth metal, and oxygen gas, and found that 
the ZnBiSiO$_{4}$ distorted spinel structure is lower in energy by about 
$0.25$~eV per formula unit. A major challenge involved in this synthesis would 
be to prevent bismuth from further oxidizing and causing the constituents to 
segregate. The conventionally determined oxidation state of Bismuth appears to 
pose a significant obstacle in synthesizing the proposed materials. However, 
that this oxidation state characterizes nearby insulating states is crucial to 
providing an odd electron formula unit and stabilizing the Dirac-semimetal 
state. Similar configurations of bismuth atoms have been achieved in the 
laboratory in the construction of bulk materials built from stacks of 
two-dimensional topological insulators.\cite{NatMater.12.422} 
We therefore propose that synthesis be conducted under reducing 
conditions (high temperature and low partial pressure of O$_2$).

Finally, we note that additional symmetry breaking may allow access to exotic 
insulating phases.  The low-energy theory at the Fermi surface can be written 
as $\cH(\bk) = v_x k_x\gamma_x+v_yk_y\gamma_y +v_zk_z\gamma_z$, centered at T, 
where $v_i$ are the Fermi velocities and $\gamma_i$ are $4 \times 4$ Dirac 
matrices. The Dirac matrices are constrained by the invariance of $\cH(\bk)$ 
under the little group at T. Orienting the $k_z$ axis to point along the line 
from T to $\Gamma$, $\cH(\bk)$ takes the form,
\begin{eqnarray}
\cH(\bk) &= v_x k_x \sigma_x\otimes\sigma_z+ v_yk_y 
(\cos\theta\sigma_x\otimes\sigma_x\\ \nonumber &+\sin\theta\sigma_x\otimes\sigma_y)+ v_z 
k_z\sigma_y\otimes \bbI. \label{kdotp}
\end{eqnarray}
Here $\theta$ is an arbitrary real parameter, $\sigma_i$ are the usual Pauli 
matrices, and $\bbI$ is the $2\times 2$ identity matrix. The exact values of 
$\theta$ and $v_i$ depend on microscopic features. 

The elements of the Little group that stabilize the Dirac point at T are: 
mirror symmetry $\cM_z$ about the $k_xk_y$ plane, sublattice (inversion) 
symmetry $\cI$, and time reversal symmetry $\Theta$. In the basis of 
Eq.~(\ref{kdotp}), these operators can be represented as, 
$\cM_z=\sigma_x\otimes \bbI$, $\cI=\sigma_z\otimes \bbI$, and $\Theta = 
i\sigma_y\otimes \bbI~K$ where $K$ denotes complex conjugation. Symmetry 
breaking perturbations lead to insulating (topological and normal) as well as 
topological semimetallic (Weyl) phases. The distorted spinel structures 
discussed in this paper, if engineered or discovered naturally, can be used to 
access such phases. Dirac semimetals are unique in that they exist at a 
multicritical point from which many exotic insulating and topological 
semimetallic phases can be reached~\cite{PhysRevLett.108.140405}.

We emphasize that the crucial feature of these materials is the network of 
interpenetrating, symmetry-related sublattices of bismuth atoms with unpaired 
electrons, a physical manifestation of the symmetry-derived result that Dirac 
points can only exist in nonsymmorphic space groups on the BZ surface.  The 
rest of the atoms of the lattice can be thought of as an insulating scaffolding 
that stabilizes this metallic bismuth network that hosts the Dirac point. The 
combination of close packing and unconventional oxidation state reduces the 
tendency towards dimerization and the system remains at the critical point of 
this half-filled state. These offer important insight into both the physics and 
materials science of the Dirac semi-metal state, and will inform efforts to 
realize such a material.
\section*{Acknowledgments} This work was supported in part by the
MRSEC program of the National Science Foundation under
Grant No. DMR11-20901 (S.M.Y.), by the Department of
Energy under Grant No. FG02-ER45118 (E.J.M. and S.Z.),
and by the National Science Foundation under Grants No.
DMR11-24696 (A.M.R.) and No. DMR09-06175 (C.L.K.). J.A.S. was supported by the REU Program at
LRSM, University of Pennsylvania. S.M.Y. acknowledges
computational support from the High Performance Computing
Modernization Office.
\makeatletter

\end{document}